# Exciton Shift Currents: DC Conduction with Sub-bandgap Photo Excitations

Y.-H. Chan[1,2,3], Diana Y. Qiu[1,2], Felipe H. da Jornada[1,2], and Steven G. Louie[1,2]

[1]**Department of Physics, University of California, Berkeley, CA, 94720-7300, USA**

[2]**Materials Sciences Division, Lawrence Berkeley National Laboratory, Berkeley, CA, 94720, USA**

[3]**Institute of Atomic and Molecular Sciences, Academia Sinica, Tapei 10617, Taiwan**

**Abstract**

Shift current is a DC current generated from nonlinear light-matter interaction in a non-centrosymmetric crystal and is considered a promising candidate for next generation photovoltaic devices. The mechanism for shift currents in real materials is, however, still not well understood, especially if electron-hole interactions are included. Here, we employ a first-principles interacting Green's-function approach on the Keldysh contour with real-time propagation to study photocurrents generated by nonlinear optical processes under continuous wave illumination in real materials and demonstrate a strong DC shift current at subbandgap excitation frequencies in monolayer GeS due to strongly bound excitons, as well as a giant excitonic enhancement in the shift current coefficients at above bandgap photon frequencies. Our results suggest that atomically thin two-dimensional materials may be promising building blocks for next generation shift-current devices

When continuous wave (CW) light is shone on a noncentrosymmetric crystal, a direct current (DC) can arise due to a second-order optical response of the crystal. The origin of this current is viewed to be related to the "shift" (1-4) of the intracell coordinates of the excited electron. This so-called shift current is proposed as an alternative to the photocurrent generated by traditional semiconductor p-n junctions for photovoltaic applications (5, 6). Unlike conventional photovoltaic devices, shift current is a bulk phenomenon, which does not require a p-n junction to separate the optically generated electron-hole pair for a DC current. Moreover, recent studies reveal that the photo-carriers in shift current can have long travel distances, which is distinct from the usual drift transport mechanism in traditional solar cells (7, 8) and makes shift current a promising candidate for efficient energy conversion.

Despite many investigations over the past decade, basic understanding of shift currents is far from complete. Most theoretical studies to date rely on the assumption of having non-interacting particles (3-6,9-11). Given that it is well-known that light-induced electron-hole pairs can form bound or resonant excitons (correlated electron-hole states), which dominate and qualitatively change the absorption features of semiconductors, excitons are expected to play a large role in shift currents, especially for reduced dimensional systems. However, it is not straight forward to generalize existing *ab initio* methods (such as the *ab initio* GW plus Bethe-Salpeter equation (GW-BSE) approach (12)), used to understand and compute excitonic effects in linear optical absorption to nonlinear optical responses. Different model approaches to investigate the effects of many-electron interactions on nonlinear optical responses of materials have been proposed. For instance, a Floquet-based model Hamiltonian formalism showed that excitonic effects enhance nonlinear response (13). In the specific case of second harmonic generation, first-principles approaches have been developed and applied to real materials, for instance, by making approximation to the full many-body perturbation theory (MBPT) treatment (14, 15) and to time-dependent density function theory where electron interaction effects are taken into account via simplified kernels (16). A real-time formulation based on propagating the time-dependent Schrodinger equation has also been developed (17) and applied to second harmonic generation (18). For shift currents, only one recent study considered the effects of excitons on the *linear* optical coefficient that might influence shift currents; but these authors included only the effects of excitons to the electromagnetic field profile in a bulk sample, and the crucial process of shift current generation itself is still treated within an independent-particle picture (11). Thus, there is still no first-principles calculation and understanding of the role of many-electron interactions, particularly those due to excitons, on shift currents.

Here, we show from first principles that 1) bound exciton states in the band gap can generate substantial shift currents, and 2) excitonic effects in the electron-hole continuum part of the spectrum can also greatly enhance shift currents due to the coherence of the electron-hole pairs.

**Methods**

To carry out the calculation of shift currents including many-electron interaction and therefore excitonic effects, we employ an *ab initio* approach to study in general nonlinear optical phenomena in real materials based on real-time propagation of the nonequilibrium interacting Green's function formalism on the Keldysh contour (19, 20). Our method allows us to consistently include many-body interactions on both linear and nonlinear optical processes, which contrasts with previous simplified first-principles study (11) that only includes the influence of excitonic effects on the profile of the light intensity as it penetrates a sample, but still treats the intrinsic shift-current response within an independent-particle picture. As we demonstrate below, the inclusion of excitonic effects on the shift-current process is critically important, since the shift-current conductivity depends on virtual transitions between excitonic states and can dramatically altered the intrinsic nonlinear response when electron-hole interactions are strong. We illustrate this physics of many-electron interactions by computing the shift-current coefficient tensor as a function of incident light frequency for a system that is atomically thin – a monolayer GeS. In addition to being the first *ab initio* calculation of shift currents from subbandgap and above-band gap frequency excitations in a real material, we show that self-energy and excitonic effects not only can enhance the total integrated shift-current density by orders of magnitude, but many spectral features of the shift-current coefficients are not even qualitatively captured by an independent particle (IP) picture. In particular, we show that, in monolayer GeS, excitonic effects strongly mix band states with different wavevector **k** in the Brillouin zone and thus yield a polarization anisotropy in the shift current that is much larger than what is found within an IP picture.

We start with the equation of motion for the interacting single-particle Green's function G on the Keldysh contour *C*,

$$\left[i\frac{d}{dt} - H(t)\right] G(t,t') = \delta(t,t') + \int_C \Sigma(t,\bar{t}) G(\bar{t},t') \, d\bar{t}, \qquad (1)$$

where $H = H_0 - e\mathbf{E}(t) \cdot \mathbf{r}$ is a mean-field Hamiltonian, which composes the electronic mean-field crystal Hamiltonian $H_0$ and an arbitrary time-dependent uniform external field **E** (which may be the field of the light); $\Sigma$ is the electron self-energy operator and *G* is the contour-ordered Green's function (see Supplemental Information). There is an equivalent adjoint equation for the time-evolution over *t'* and we note that the $\mathbf{r}$ operator needs to be treated with care for a crystal. The self-energy is computed within the GW approximation, i.e., $\Sigma=iGW$, where *W* is the screened Coulomb potential. (21) The solution to Eq. 1 provides the means to calculate various physical quantities, including the shift current.

One challenge in solving Eq. 1 is that the time evolution of *G* over *t* and *t'* has to be performed simultaneously. Following Ref. (22), we decouple these equations by splitting the self-energy into that at equilibrium, $\Sigma[G_0]$, plus a correction term, $\delta\Sigma[G,G_0] = \Sigma[G] - \Sigma[G_0]$, where $G_0$ is the equilibrium Green's function with no external field. We evaluate $\Sigma[G_0]$ within the GW approximation ($\Sigma^{GW}$), and the correction $\delta\Sigma[G,G_0]$ within a static-screening scheme (*i.e.*, the so-called static COHSEX) approximation. (21) In the weak field limit, it has been shown that the resulting linearized equation of motion for *G* obtained this way yields an optical spectrum identical to the one from the *ab initio* GW-BSE approach computed with a statically screened electron-hole interaction kernel (22, 23, 24). Since this approach captures quasiparticle excitation given by the fully dynamical GW approximation at equilibrium and reproduces the optical response of GW-BSE near equilibrium by neglecting memory effects, we refer to it as the *time-dependent adiabatic GW* (TD-aGW) approach in analogy to an adiabatic approximation to time-dependent density functional theory.

Within the TD-aGW approach, the equation of motion of *G* over *t* and *t'* can be rigorously rewritten (22) in terms of the interacting single-particle density matrix $\rho$, evaluated in a quasiparticle basis $\rho_{nm,\mathbf{k}} \equiv \langle n\mathbf{k}|\rho|m\mathbf{k}\rangle$,

$$i\hbar\frac{\partial}{\partial t}\rho_{nm,\mathbf{k}}(t) = [H_0(t) + \Sigma^{GW} + \delta\Sigma(t) - e\mathbf{E} \cdot \mathbf{r}, \rho]_{nm,\mathbf{k}}, \qquad (2)$$

where *n* and *m* are band indices, and **k** is a **k** point in the Brillouin zone.

A challenge in dealing with Eq. 2 is that the position operator $\mathbf{r}$ is difficult to handle for extended states, although it can be formally separated into well-defined interband ($\mathbf{r}^{inter}$) and intraband ($\mathbf{r}^{intra}$) parts (25). Unlike in the case of linear optical processes in semiconductors at equilibrium and low temperature for which only the interband part of $\mathbf{r}$ contributes, we also need to evaluate here the commutator $[\mathbf{r}^{intra},\rho]_{nmk}$, which involves the derivative of the density matrix with respect to different **k** points, which in turn involves an arbitrary **k**-dependent gauge. Here, we introduce the use of a *locally smooth gauge* using an idea similar to the covariant derivative (26, 27): we rotate wave functions at nearby **k** points so that the overlap of connected wave functions is Hermitian (26, 28). We find this local smooth gauge to give equivalent results to a global smooth gauge but at a negligible computational cost (see Supplemental Information). This technique is also an efficient alternative to replacing the general derivative by a summation of velocity and position matrices, as is commonly done in the field (4, 29).

Once the density matrix is known as a function of time through a real time propagation of Eq. 2, we may compute physical observables, such as the polarization and current density, by taking traces of the product of an appropriate operator with the density matrix. We compute the time-dependent current density as $\boldsymbol{J}(t) = Tr(\rho(t)\mathbf{v})$, where **v** is the velocity operator. To extract the shift current conductivity tensor, we evaluate the discrete Fourier components $J_n$ from the current density computed under the coupling to a CW monochromatic field (16) $\boldsymbol{E}(t) = \boldsymbol{E}_0 \sin(\omega_E t)$. Responses at different harmonic frequencies can be computed by fitting to a linear equation $\boldsymbol{J}(t) = \sum_{n=-S}^{S} \boldsymbol{J}_n e^{-i\omega_n t}$, where $\omega_n = n\omega_E$ with n an integer and $S$ denotes the cutoff component of the Fourier series. The linear conductivity at finite frequencies, and hence the absorption coefficient, can be computed from $\boldsymbol{J}_1$. The shift current density is obtained from $\boldsymbol{J}_0$, and the second harmonic generation (SHG) tensor can be computed from $\boldsymbol{J}_2$. More generally, higher harmonic responses may also be obtained from the fitting to Fourier components beyond $\boldsymbol{J}_2$.

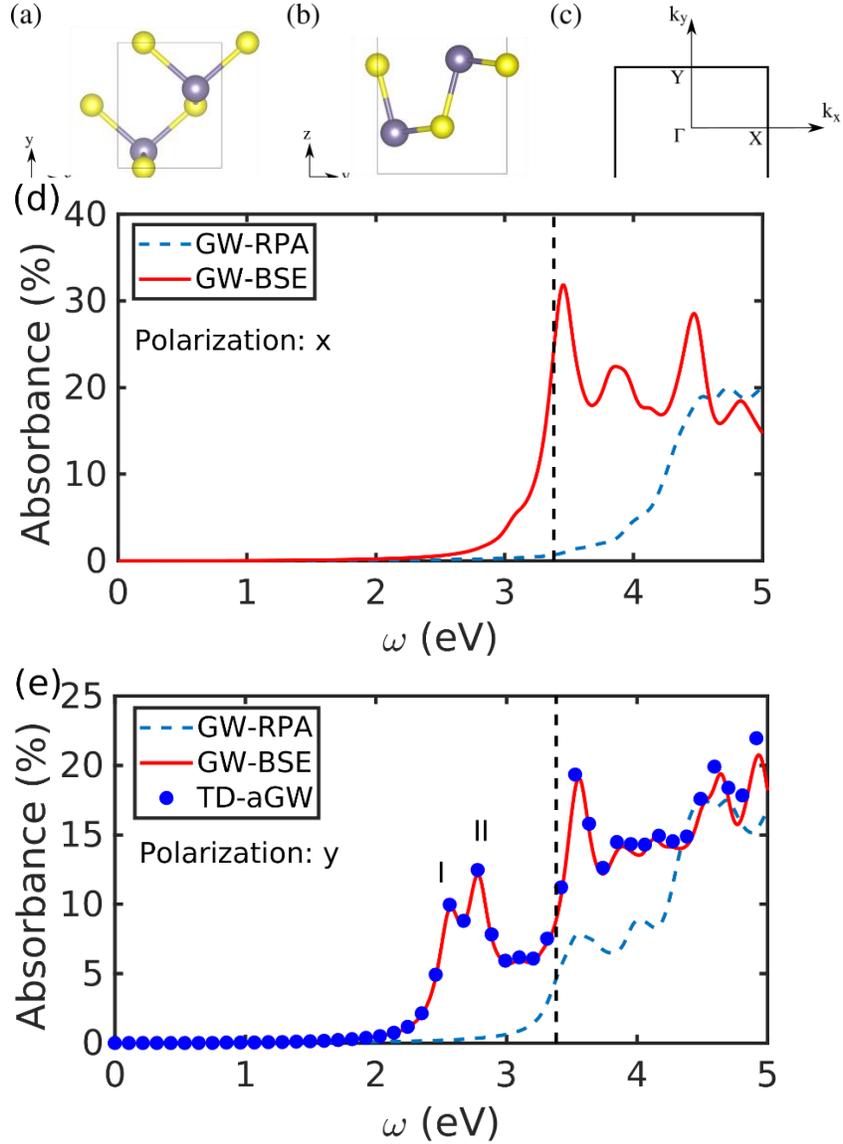

Fig. 1: Atomic model of a unit cell of monolayer GeS, (a) top view and (b) side view. (c) The Brillouin zone of monolayer GeS. Calculated linear absorption spectrum of monolayer GeS with x-polarized (d) and y-polarized (e) light at normal incidence: independent-particle approximation with GW quasiparticle energies (GW-RPA, blue dashed line), GW-BSE calculation (GW-BSE, red line), and TD-aGW calculation (TD-aGW, blue dots and only y-polarization results are shown). Both GW-BSE and TD-aGW results include electron-hole interaction (excitonic) effects, and all curves are generated with a Lorentzian broadening factor of 100 meV. Vertical dashed lines in (d) and (e) indicate the equilibrium band gap. Peaks I and II in (e) correspond to absorption from formation of bound excitons within the bandgap.

## Results and Discussion

We now present our results on excitonic effects on the shift current response tensor in monolayer GeS in the structure shown in Fig. 1 (a) and (b), which was reported to have a relatively large shift current response in the IP approximation (30). We first verify the linear absorption spectrum calculated within our TD-aGW method against that from the *ab initio* GW-BSE approach, which was performed using the BerkeleyGW package (12, 21, 31). Density-functional theory (DFT) calculations (as the starting mean-field for the GW-BSE and TD-aGW calculations) were performed using the Quantum Espresso package (32). (see SI for computational details). The direct band gap for monolayer GeS as computed from DFT and GW are 1.90 eV and 3.38 eV, respectively.

Figure 1 depicts the computed linear absorption spectrum of monolayer GeS with normal incidence light polarized either along the zigzag (Fig. 1 (d)) or armchair direction (Fig. 1(e)), called the x and y directions, respectively. The GW-RPA (IP approximation with GW quasiparticle energies) spectrum has an onset at 3.4 eV corresponding to the quasiparticle direct band gap from our GW calculation. Comparing with the spectrum from the GW-BSE calculation, we find a binding energy of near 1.0 eV for the lowest energy optically active exciton, which agrees well with previous studies (33, 34).

We validate our TD-aGW method and code by comparing the linear absorption results from TD-aGW with those from the GW-BSE calculations. To obtain the linear responses at all frequencies, we apply an electric impulse at the beginning of the simulation and evaluate the time-dependent electric polarization $P^e(t)$. The absorption spectrum $\epsilon_2(\omega)$ is computed from $\epsilon_2^{ab}(\omega) = I[P^{e,a}(\omega)/(\epsilon_0 E^b(\omega))]$, where $a$ and $b$ denote different components of the dielectric tensor, relating to the direction of the electric polarization and electric field. Our TD-aGW results are shown in Fig. 1 (e) by the blue dots, and are virtually identical to the GW-BSE results.

In the time-propagation calculations, the contribution of light with a specific frequency ω to the DC shift current density along the *a* direction, denoted by $J_0^a(\omega)$, is given by (4)

$$J_0^a(\omega) = 2 \sum_{bc} \sigma^{abc}(0;\omega,-\omega)E^b(\omega)E^c(-\omega), \qquad (3)$$

where $\sigma^{abc}$ is the second-order conductivity tensor, and *a*, *b*, and *c* are indices indicating the component of the shift current density and the polarization of the light field ($E^b$ and $E^c$), in Cartesian coordinates, respectively. The second-order conductivity tensor in principle can also be computed without performing an explicit time-evolution of the density matrix, but it is quite conceptually and computationally involved for an interacting many-body system. Within the independent-particle (IP) approximation, however it can be obtained from the mean-field band structure and optical transition matrix elements through the so-called sum-over-band formula (4). It is within this formulation that it has been shown in recent studies (4, 35) that the shift current in the IP approximation depends on the absolute value of the interband transition matrix elements and the shift vectors, which describe the change of the intracell position of Bloch wavefunction between the valence band and conduction band involved in the optical transition.

In Fig. 2, we present two components of the shift current conductivity tensor as a function of photon frequency, with and without electron-hole coupling included in the calculations. The calculations were done on a 24×24×1 **k**-point mesh, and, for physical lifetime and convergence reasons, a dephasing factor that is equivalent to a spectral Lorentzian broadening of 100 meV was use in the time propagation. First, we see the striking results that there are large DC currents generated by *subbandgap photon frequencies* at the exciton excitation energies. This remarkable phenomenon arises from many-body (electron-hole) interaction effects. Second, we see that, at frequencies above the band gap around 3.5 eV, the largest peak in the *yyy* component and in the *yxx* component are both enhanced by *several orders of magnitude* as compared to the IP curves. This clearly illustrates the dramatic effects of excitonic effects on shift current generation even inside the two-particle continuum. The enhancement factors here are spectacularly larger than those in the linear absorption where excitonic effects at frequencies above the band gap typically give a factor of two enhancement. Upon including excitonic effects, the total shift current density integrated over the solar spectrum for the yyy components increases from 0.2 A/m² to 4.5 A/m², while that for the yxx components increases from 0.5 A/m² to 8.4 A/m² in monolayer GeS – roughly a 20 times enhancement compared with the non-interacting cases.

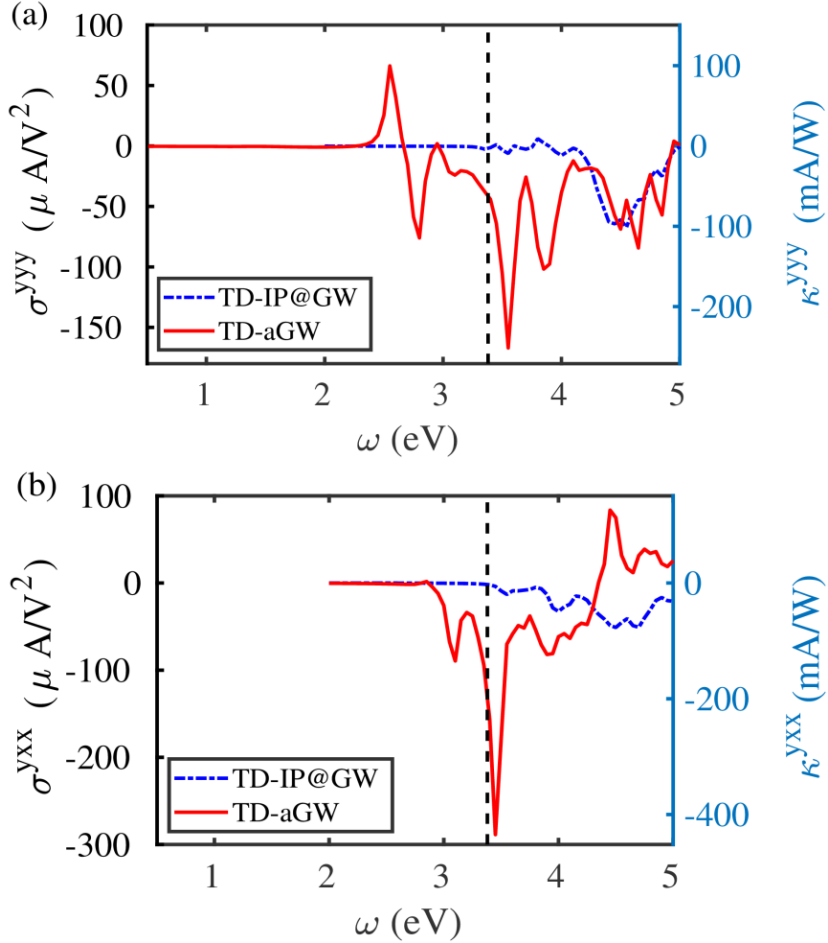

Fig. 2: Comparison of the shift current tensor components as function of frequency of monolayer GeS computed within a time-dependent independent-particle (TD-IP) formalism with quasiparticle energies obtained from a GW calculation (blue dash-dotted line) and the TD-aGW (red solid line) method for (a) *yyy* and (b) *yxx* tensor components with the layer thickness taken to be 2.6 Å. Black dashed line indicates the position of the direct quasiparticle bandgap. The right axis shows the corresponding response converted to units of photoresponsivity defined as the generated current per unit of incident radiant power.

To get a better understanding of excitonic effects on the shift current responses, it is useful to formally look at an approximated sum-over-band formula derived from a linearized equation of motion (36). The approximated shift current tensor for a two-band model (with exciton states label by m and n) is

$$\sigma^{abc}(0;\omega,-\omega) \sim \frac{-e^3}{m_e V_{\text{xtal}}} \sum_{m,n} \frac{P_m^a Q_{mn}^b \Omega_n^c}{E_m(E_n - \hbar\omega - i\eta)}, \qquad (4)$$

where $n$, $m$ are indices of the exciton states with a **k**-space envelope function of the $n$th exciton wavefunction given by $\phi_{cvk}^n$ and energy $E_n$ (i.e., the many-body exciton state is expressed as $|\psi_m\rangle = \sum_{cvk} \phi_{cvk}^m |cvk\rangle$ with $|cvk\rangle$ an free electron-hole interband excitation from a valence band $v$ to a conduction band $c$ at the wavevector **k**), $m_e$ is the electron mass, $V_{xtal}$ is the crystal volume, $P_m = \sum_k \phi_{cvk}^{(m)} p_{vck}$ is an exciton transition matrix element, $\Omega_n = \sum_k \phi_{cvk}^n r_{cvk}^{inter}$ is an exciton position matrix element, and $Q_{mn} = i\langle m|\partial_k|n\rangle$ is an inter-exciton position matrix element. We can rewrite $P_m = \langle 0|p|m\rangle$ and $\Omega_n = \langle 0|r^{inter}|n\rangle$, which are the corresponding transition amplitudes between the ground state $|0\rangle$ and an exciton state $|m\rangle$ or $|n\rangle$.

From Eq. (4), we see that the existence of bound excitons leads to an in-gap shift current response. In practice, the shift current tensor in the subbandgap frequency regime will be broadened due to inhomogeneous broadening and finite exciton lifetime effects, which are different from the traditional electron scattering-time effects in the Drude conductivity. Moreover, the shift current is characteristically different from linear optical absorption. While the absorption is determined by $|\langle 0|r^{inter}|n\rangle|^2$ summed over exciton states, shift current involves sum over the product of distinct optical transition matrices of transition from the ground state to two exciton states (m and n in Eq. (4)) that may be different. Hence, the sharp exciton features in the shift-current conductivity may exhibit strong optical anisotropy depending on the incoming light polarization. We find that the first peak of the *yyy* tensor component of the shift current for monolayer GeS at 2.5 eV corresponds to the first peak in the linear absorption at the same energy, and the second feature (a dip) in the response corresponds to the second peak in the linear absorption at 2.8 eV which is due to an exciton in the valley along Γ–Y direction in the Brillouin zone. At other energies, the spectral features in the shift current conductivity tensor appear less sharp than the corresponding sharp excitonic peaks in the linear absorption spectrum.

We now elucidate the physics of the excitonic enhancement of the shift current in the two-particle (electron-hole) continuum. In the IP case, the spectral features in the shift-current conductivity correlates with features in the shift vectors $R_{cvk} \equiv \partial_k \varphi_{cvk} + r_{cck} - r_{vvk}$, where the phases $\varphi_{cvk}$ are defined as $r_{cvk} = |r_{cvk}|e^{-i\varphi_{cvk}}$. As shown above, we can define analogous inter-exciton transition matrix elements $Q_{nm}$ in the excitonic case. However, the magnitudes of $Q_{mn}$ do not significantly increase compared to the non-interacting case, since $Q_{mn}$ can be regarded as a linear combination of the shift vectors (see SI). Instead, the enhancement of the shift-current conductivity comes primarily from an enhancement in the optical transition matrix elements $P_m^a$ and $\Omega_n^c$, in particular if excitons *m* and *n* are strongly bound in Eq. (4). For example, in the case of extremely tightly bound excitons, the corresponding exciton envelope wavefunctions are extended over the Brillouin zone, resulting in $\phi_k \sim 1/\sqrt{V_{xtal}}$, so $P_m^a \Omega_n^c \sim V_{xtal}$ and thus a finite number of bound excitons give a finite contribution to the shift current conductivity, instead of a vanishing contribution for a single vertical transition at a single k point in the IP case. This excitonic enhancement of the shift current conductivity due to an enhancement of the optical dipole matrix elements is analogous to the case of linear optical absorption.

We now discuss how the presence of bound excitons affects the photovoltaic power efficiency of shift-current materials. The power-conversion efficiency is related to both the short-circuit current and the open-circuit voltage. We have already shown that the current response (i.e., the short-circuit current) gets dramatically enhanced due to excitonic effects. For instance, comparing with the bulk ferroelectric material BaTiO$_3$, which has a shift-current conductivity response of 5 μA/V$^2$ (5) in the same frequency range, monolayer GeS displays a response that is more than one order of magnitude larger and inside the electronic bandgap. It is of the same order of magnitude as the recently reported response in the Weyl semimetal TaAs. The photoresponsivity $\kappa$ (defined as the generated current per unit of incident radiant power; see $\kappa$ on the right axis of Fig. 2) of monolayer GeS is also comparable to Si-based solar cells with reported $\kappa$=400 mA/W. (37) However, the total current density output by monolayer GeS when illuminated under the solar spectrum is at most only 4.5 A/m$^2$ (see Supplementary Information), roughly two orders of magnitude smaller than that in bulk Si. This smaller total current is due to the optical response of monolayer GeS not overlapping well with the solar spectrum and being restricted to a small range at high frequencies owing to its large band gap. We note that the photoresponsivity $\kappa$ of Si here refers to a solar cell which typically has a Si thickness of many microns to capture all the incoming photons, while the $\kappa$ of a monolayer GeS device here is for an atomically thin material of thickness of ~0.2 nm and may be significantly increased by incorporating the GeS into a layered tandem device. Other materials with smaller band gaps, such as GeSe, GeTe, SnS, SnSe, and SnTe, as suggested by previous first-principles calculations (30), might be better able to fully harness the solar spectrum using shift currents for photovoltaic applications.

While the short-circuit current depends on the shift current conductivity and not sensitively on the details of the sample, the same is not true for the open-circuit voltage. For instance, previous studies (38) argued that the open-circuit field $E_{OC}$ for a shift-current device would increase for samples with shorter carrier lifetimes (*e.g.*, in disordered samples), as generated carriers cannot easily travel through the material. However, this previous conclusion assumes that the free carries present in the material are photo-induced and not coming from ionized defects. In a real material such as GeS, however, we expect that, even for clean samples, having a bulk DC conductivity of $\sigma^{DC} \sim 10^{-2}(\Omega m)^{-1}$, the number of free carriers produced by ionizing defects will be of the order of $10^6 cm^{-2}$, roughly four order of magnitude larger than the photogenerated ones at steady state (see Supplementary Information). Therefore, the open-circuit field will clearly depend on the sample quality and defect concentration.

We can estimate the open-circuit field in the case of a finite concentration of shallow defects as $E_{oc} = J_{sc}/\sigma^{dc}$, where $\sigma^{dc}$ is the sample's conventional DC conductivity. Using a current density of 4.5 A/m² for GeS under sun light and a small bulk conductivity $10^{-2}(\Omega m)^{-1}$, we estimate the open-circuit field to be 450 V/m. From these, we see that the open-circuit voltage depends on the sample quality and the geometry of the device but *not* directly on the band gap of the material, which is in contrast to conventional solar cell devices based on traditional p-n junction. Consequently, it would be possible to take advantage of the excitonic enhancement of short-circuit current and also obtain a large open-circuit voltage by minimizing the number of shallow impurity defects in shift-current materials.

In summary, we have developed a theoretical framework and computer code to study nonlinear optical phenomena in real materials from *ab initio* including excitonic effects, and used the approach to calculate the shift current of monolayer GeS. We have shown that excitonic effects on the shift current give rise to DC current from optical excitations with in-gap frequencies, as well as dramatic enhancement for above-gap frequencies. For monolayer GeS, the total current density integrated over solar spectrum shows a 20 times enhancement due to exciton effects as compared to the free electron-hole case. These findings add promise for using shift current for photovoltaic applications, allowing greater tunability and overlap with the solar spectrum. This work reveals the central importance of excitonic effects in shift current generation and open new pathways for designing highly-efficient shift current devices from quasi-2D materials. It provides a new perspective on the search for shift current materials for applications with efficiencies not bound by the Shockley-Queisser limit in traditional single p-n junction devices.

**Acknowledgments.** This work was supported by the Center for Computational Study of Excited State Phenomena in Energy Materials (C2SEPEM), which is funded by the U.S. Department of Energy, Office of Science, Basic Energy Sciences, Materials Sciences and Engineering Division under Contract No. DE-AC02-05CH11231, as part of the Computational Materials Sciences Program. YHC thanks C. Attaccalite, T. Rangel and T. Morimoto for helpful discussion. We acknowledge the use of computational resources at the National Energy Research Scientific Computing Center (NERSC), a DOE Office of Science User Facility supported by the Office of Science of the U.S. Department of Energy under Contract No. DE-AC02-05CH11231.

# References

[1] Kraut W, von Baltz R (1979) Anomalous bulk photovoltaic effect in ferroelectrics: A quadratic response theory. Phys. Rev. B 19(3):1548–1554.
[2] Sturman B. I., Fridkin, V. M. (1992) The photovoltaic and photorefractive effects in noncentrosymmetric materials. (Gordon and Breach Science Publishers)

[3] Aversa C, Sipe JE (1995) Nonlinear optical susceptibilities of semiconductors: Results with a length-gauge analysis. Physical Review B 52(20):14636–14645.
[4] Sipe JE, Shkrebtii AI (2000) Second-order optical response in semiconductors. Physical Review B 61(8):5337–5352.
[5] Tan LZ, et al. (2016) Shift current bulk photovoltaic effect in polar materials–hybrid and oxide perovskites and beyond. Npj Computational Materials 2:16026.
[6] Cook AM, Fregoso BM, Juan Fd, Coh S, Moore JE (2017) Design principles for shift current photovoltaics. Nature Communications 8:14176.
[7] Nakamura M, et al. (2017) Shift current photovoltaic effect in a ferroelectric charge-transfer complex. Nature Communications 8(1):281.
[8] Ogawa N, Sotome M, Kaneko Y, Ogino M, Tokura Y (2017) Shift current in the ferroelectric semiconductor SbSI. Phys. Rev. B 96(24):241203.
[9] Ibanez-Azpiroz J, Tsirkin S. S., and Souza I. (2018) *Ab initio* calculation of the shift photocurrent by Wannier interpolation. Phys. Rev. B 97(24): 245143
[10] Wang C, Liu X, Kang L, Gu B. L., Xu Y., and Duan W. (2017) First-principles calculation of nonlinear optical responses by Wannier interpolation. Phys. Rev. B 96(11): 115147
[11] Fei R, Tan L, Rappe A. (2020) Shift current bulk photovoltaic effect influenced by quasiparticles and excitons. Phys. Rev. B 101:04510.
[12] Rohlfing M, Louie SG (2000) Electron-hole excitations and optical spectra from first principles. Phys. Rev. B 62(8):4927–4944.
[13] Morimoto T, Nagaosa N (2016) Topological aspects of nonlinear excitonic processes in noncentrosymmetric crystals. Phys. Rev. B 94(3):035117.
[14] Chang EK, Shirley EL, Levine ZH (2001) Excitonic effects on optical second-harmonic polarizabilities of semiconductors. Physical Review B 65(3).
[15] Leitsmann R, Schmidt WG, Hahn PH, Bechstedt F (2005) Second-harmonic polarizability including electron-hole attraction from band-structure theory. Physical Review B 71(19).
[16] Luppi E, Hbener H, Vniard V (2010) *Ab initio* second-order nonlinear optics in solids: second-harmonic generation spectroscopy from time-dependent density-functional theory. Physical Review B 82(23):235201.
[17] Attaccalite C, Grüning M (2013) Nonlinear optics from an *ab initio* approach by means of the dynamical berry phase: Application to second- and third-harmonic generation in semiconductors. Phys. Rev. B 88(23):235113.
[18] Grüning M, Attaccalite C (2014) Second harmonic generation in h-BN and MoS$_2$ monolayers: Role of electron-hole interaction. Phys. Rev. B 89(8):081102.
[19] L.V.Keldysh (1965) Diagram technique for nonequilibrium processes. Sov. Phys. JETP 20(4):1018.
[20] Kadanoff LP, Baym G (1962) Quantum Statistical Mechanics.
[21] Hybertsen MS, Louie SG (1986) Electron correlation in semiconductors and insulators: Band gaps and quasiparticle energies. Phys. Rev. B 34(8):5390–5413.
[22] Attaccalite C, Grüning M, Marini A (2011) Real-time approach to the optical properties of solids and nanostructures: Time-dependent Bethe-Salpeter equation. Phys. Rev. B 84(24):245110.
[23] Rocca D, Lu D, Galli G (2010) Ab initio calculations of optical absorption spectra: Solution of the Bethe–Salpeter equation within density matrix perturbation theory. The Journal of Chemical Physics 133(16):164109.
[24] Rabani E, Baer R, Neuhauser D (2015) Time-dependent stochastic Bethe-Salpeter approach. Physical Review B 91(23):235302.
[25] Blount EI (1962) Formalisms of Band Theory. Vol. 13, pp. 305–373.
[26] Souza I, Íñiguez J, Vanderbilt D (2004) Dynamics of berry-phase polarization in time-dependent electric fields. Phys. Rev. B 69(8):085106.
[27] Virk KS, Sipe JE (2007) Semiconductor optics in length gauge: A general numerical approach. Phys. Rev. B 76(3):035213.
[28] Marzari N, Vanderbilt D (1997) Maximally localized generalized wannier functions for composite energy bands. Phys. Rev. B 56(20):12847–12865.

[29] Cabellos JL, Mendoza BS, Escobar MA, Nastos F, Sipe JE (2009) Effects of nonlocality on second-harmonic generation in bulk semiconductors. Phys. Rev. B 80(15):155205.
[30] Rangel T, et al. (2017) Large bulk photovoltaic effect and spontaneous polarization of single-layer monochalcogenides. Phys. Rev. Lett. 119(6):067402.
[31] Deslippe J, et al. (2012) BerkeleyGW: A massively parallel computer package for the calculation of the quasiparticle and optical properties of materials and nanostructures. Computer Physics Communications 183(6):1269–1289.
[32] Giannozzi P, et al. (2009) Quantum Espresso: a modular and open-source software project for quantum simulations of materials. Journal of Physics: Condensed Matter 21(39):395502.
[33] Gomes LC, Trevisanutto PE, Carvalho A, Rodin AS, Castro Neto AH (2016) Strongly bound mott-wannier excitons in GeS and GeSe monolayers. Phys. Rev. B 94(15):155428.
[34] Xu L, Yang M, Wang SJ, Feng YP (2017) Electronic and optical properties of the monolayer group-IV monochalcogenides MX (M = Ge, Sn; X = S, Se, Te). Phys. Rev. B 95(23):235434.
[35] Morimoto T, Nagaosa N (2016) Topological nature of nonlinear optical effects in solids. Science Advances 2(5): e1501524.
[36] Pedersen TG (2015) Intraband effects in excitonic second-harmonic generation. Phys. Rev. B 92(23):235432.
[37] Gavin B. Oterhoudt, et al. (2019) Colossal mid-infrared bulk photovoltaic effect in a type-I Weyl semimetal. Nature Materials. 10.1038
[38] Morimoto T, Nakamura M, Kawasaki M, Nagaosa N (2018) Current-voltage characteristic and shot noise of shift current photovoltaics. Phys. Rev. Lett. 121(26):267401.
[39] Bradley AJ, et al. (2015) Probing the role of interlayer coupling and coulomb interactions on electronic structure in few-layer MoSe$_2$ nanostructures. Nano Letters 15(4):2594–2599. PMID: 25775022.
[40] Qiu DY, da Jornada FH, Louie SG (2017) Environmental screening effects in 2d materials: Renormalization of the bandgap, electronic structure, and optical spectra of few-layer black phosphorus. Nano Letters 17(8):4706–4712. PMID: 28677398.
[41] Perdew JP, Burke K, Ernzerhof M (1996) Generalized gradient approximation made simple. Phys. Rev. Lett. 77(18):3865–3868.
[42] Hamann DR (2013) Optimized norm-conserving Vanderbilt pseudopotentials. Phys. Rev. B 88(8):085117.
[43] Schlipf M, Gygi F (2015) Optimization algorithm for the generation of ONCV pseudopotentials. Computer Physics Communications 196:36–44.
[44] da Jornada FH, Qiu DY, Louie SG (2017) Nonuniform sampling schemes of the Brillouin zone for many-electron perturbation-theory calculations in reduced dimensionality. Phys. Rev. B 95(3):035109.
[45] ASTM Standard G173. (2008). Standard tables for reference solar spectral irradiances: direct normal and hemispherical on 37° tilted surface. West Conshohocken, PA: American Society for Testing and Materials.

# Supporting Information

## Parallel-transport gauge

Difficulty arises when numerically integrating Eq. 2 in the main text without an analytic expression for the electron wavefunctions. Due to the presence of arbitrary phases at different **k** point in the solution for the Bloch states, derivatives with respective to **k** in the intraband coupling term could give random values if care is not taken. Earlier work suggests treating this problem by using a global parallel transport gauge. (26) Here, we find that for the nonlinear responses of interest, it is sufficient to use a locally smooth gauge. We use an idea similar to the covariant derivative (26, 27, 28), which is essentially equivalent to taking a conventional derivative in the parallel transport gauge. In this method, wavefunctions at nearby **k** points are rotated so that the overlap of connected wavefunctions is Hermitian. (27, 28) The procedure is the following. We first compute the overlap $S^{k,k+\delta k}$ between a wavefunction at **k** and its neighbors. We identify wavefunctions at neighboring **k** points with large overlaps as connected and zero out the matrix elements between disconnected points. We perform a singular-value decomposition (SVD) of $S$,

$$S^{k,k+\delta k} = U\lambda V^\dagger, \tag{1}$$

And a gauge rotation matrix is constructed by

$$M = VU^\dagger. \tag{2}$$

Applying the above gauge rotation to wavefunction at $k + \delta k$ amounts to a gauge choice in which the overlap of connected wavefunctions is Hermitian. (28) In this gauge, the position operator can be also computed by taking the covariant derivative of the wavefunction, which provides an alternative of using $p \approx mv$ as is often done in first-principles calculations.

## Computational details

### GW-BSE calculations

DFT calculations are performed with the Quantum Espresso package (32). We use PBE pseudopotentials (41) from the SG15 ONCV potentials database (42, 43). For the DFT ground-state calculation of monolayer GeS with a supercell geometry, we use a **k**-mesh of 12×12×1 and a plane wave energy cutoff of 80 Ry. A vacuum of 15 Å is chosen to prevent spurious interactions between periodic images. The GW-BSE excited-state calculation is done with the BerkeleyGW package (12, 21, 31). A **k**-point grid of 24×24×1 with a subsampling of 10 points in the mini-Brillouin zone (42), a dielectric energy cutoff of 10 Ry, and 5000 bands are used in the GW calculations. The dielectric matrix is computed using the Hybertsen-Louie generalized plasmon pole model (21). The direct band gap in our DFT calculation is 1.9 eV, and the GW direct band gap is 3.38 eV.

### Real-time propagation calculations

In the TD-aGW or the TD-IP calculations, the time-integration is performed with a time step of 0.05 fs using a fourth order Runge-Kutta method. We keep 6 conduction bands and 6 valence bands in the simulation. A small broadening factor γ of 100 meV is added via a dephasing term in the Hamiltonian $-i\gamma \rho_{nmk}$ with $n \neq m$. This is equivalent to replacing a delta function with a Lorentzian function in usual absorption spectrum calculations. The total simulation time is ensured to be long enough for the system to reach a steady state. For a small broadening, we evolve the system up to 160 fs. In the TD-aGW calculation, the exchange and direct kernel are computed within the Tamm-Dancoff approximation using the BerkeleyGW code (12, 21, 31).

## Shift current from solar spectrum

The current reported in the main text for monolayer GeS is computed using the solar spectrum from the reference air mass 1.5 spectrum (45). The calculation is done by converting the shift current tensor into units of mA/W, multiplying it by the solar spectrum, and integrate the resulting current spectral density. For simplicity, we assume that the incoming sunlight is first polarized along the $y$ direction by halving the incoming solar irradiance.

## Shift vector and the excitonic enhancement

We now detail how $Q_{nm}$ can be viewed as generalized shift vectors and how the excitonic enhancement of the shift current can be seen from the enhancement of optical conductivity. In the IP picture, the shift vector contains most features of the shift current conductivity and connects the linear absorption and the shift current conductivity. However, an analogous relation can only be seen in Eq. (4) by considering the intra-exciton (n=m) transition matrix, $Q_{nn} = \sum_k |\phi_{cvk}^{(n)}|^2 (\partial_k \theta_{cvk}^{(n)} + r_{cck} - r_{vvk})$, with $\phi_{cvk}^{(n)} = |\phi^{(n)}{}_{cvk}| e^{-i\theta_{cvk}^{(n)}}$. (Here we use the exciton envelope function in **k**-space as defined in the main text.) The phase factor of the position matrix element in the shift vector $R_{cvk}$ is now replaced by the phase factor of the exciton envelope wave function modulated by the exciton wave function amplitude squared. Therefore, the physical meaning of $Q_{nn}$ is an average shift weighted by the exciton envelope wave functions, and the off-diagonal $Q_{nm}$ is a generalized shift due to inter-exciton transitions. We find for a simple tight-binding model that the spectrum weight $\sum_m Q_{nm} \delta(\omega - E_n)$ is of the same order of magnitude as $\int d\mathbf{k}\, R_{nmk} \delta(\omega - E_{cvk})$, where $E_n$ is the n-th exciton energy and $E_{cvk}$ is the transition energy from the valence band c at a k-point k.

Next, we detail how the enhancement of the shift current conductivity is achieved for strongly correlated electron-hole pairs in the excitons. The mechanism is analogous to the excitonic enhancement of the oscillator strength in the linear optical absorption. Assuming an extremely tight-bound exciton with envelope wave function $\phi_{cvk}^{(m)} = 1/\sqrt{V_{xtal}}$, as discussed in the main text, one can show for a particular transition at a single k point that the linear optical absorption in the IP picture scales as $\frac{1}{V_{xtal}} |r_{cvk}|^2$, which vanishes as the crystal volume goes to infinity (but of course remains a finite value upon integration over the Brillouin zone for a specific pair of bands). However, when excitonic effects are included, the linear optical absorption spectrum for a single bound excitonic state scales as $\frac{1}{V_{xtal}} |\sum_k \phi_{cvk} r_{cvk}|^2 = \frac{1}{V_{xtal}^2} |\sum_k r_{cvk}|^2$, which remains a finite constant as the crystal volume goes to infinity. Thus, while the contribution from a transition at a single k-point is vanishing in the IP case, a finite contribution is obtained for a single exciton due to constructive interferences. For the shift current, we can show similarly that the factors that go in the shift current conductivity for the IP and the excitonic picture are $\frac{1}{V_{xtal}} |r_{cvk}|^2 R_{cvk}$ and $\frac{1}{V_{xtal}} \sum_n \sum_k r_{cvk} \phi_{cvk}^{(n)} \sum_{k'} \phi_{cvk'}^{(n)*} (\partial_{k'} \phi_{cvk'}^{(m)} - i\phi_{cvk'}^{(m)} (r_{cck'} - r_{vvk'})) \sum_{k''} \phi_{cvk''}^{(m)*} r_{cvk''}$, respectively. For an extremely tight-bound exciton this simplifies to $\frac{1}{V_{xtal}^3} \sum_k r_{cvk} \sum_{k'} (r_{cck'} - r_{vvk'}) \sum_{k''} r_{cvk''}$, from which we conclude that the same enhancement as in the linear optical conductivity also applies. In more general cases, although the constructive interference is not perfect, we also see additional contribution from the inter-exciton transition as long as the transition matrix $Q_{nm}$ is nonvanishing. Therefore, we can expect to see at least a similar enhancement factor for the shift-current conductivity as in the case of linear optical absorption including contributions from excitons.

## Open-circuit field and photo-induced charge density

In steady-state, the open circuit field $E_{oc}$ is such that the shift current density balances the current density generated by $E_{oc}$, $-J_{sc} = J_{oc} = \sigma^{dc} E_{oc}$ (37). The shift-current density generated by $E_{ext}(\omega)$ is given by J$_{sc}$ =

$\sigma^{sc}(0; \omega, -\omega)E_{ext}(\omega)E_{ext}(-\omega)$. Next, we investigate the number of free carriers that can contribute to the open-circuit voltage, generated by ionizing impurity/defect states or by photo-induced intrinsic carriers. We write the conventional linear DC conductivity as

$$\sigma^{dc} = ne\mu$$

where $e$ is the elementary charge, $n$ is the three-dimensional carrier density, and $\mu$ is the carrier mobility. In general, besides the thermal activated carriers, there are two sources of additional carriers when a sample is illuminated under sunlight: ionized shallow defect centers and photo-induced bulk carriers across the bandgap. As we argue in the next paragraph, the contribution from the latter is negligible compared to the former.

First, we can estimate the number of excitons present in the sample when it is illuminated under sunlight within a rate-equation approach,

$$\frac{dn_{xct}}{dt} = R_{photon} - \gamma n_{xct} = 0,$$

where $n_{xct}$ is the two-dimensional exciton density, $R_{photon}$ is the rate for generating an exciton in monolayer GeS under sunlight illumination, and $\gamma$ is the relaxation rate of excitons back to the ground state. For simplicity, we take $\gamma = 50 meV$, which is a typical linewidth associated with bound excitons in monolayer materials. By using the optical absorbance of monolayer GeS, the absorbed sun light power density integrated up to 2.6 eV is about 30 W/m², which converts to $R_{photon} = 7 \times 10^{15}(cm^2 s)^{-1}$. From this photon generation rate, we obtain $n_{xct} = 42 cm^{-2}$. Even if we assume that the excitons are instantaneously thermalized and/or separated, we obtain a photo-induced charge density at steady state of $n \sim 100 cm^{-2}$, a few orders of magnitude smaller than the observed carrier density in very clean quasi-2D samples. Hence, the open-circuit voltage for a shift-current device, under these conditions, will be dictated by the DC conductivity arising from carriers due to ionized shallow impurity/defect states.